\documentstyle[11pt,aasms4]{article}

\lefthead{Horvath}
\righthead{GRBs and HI supershells}

\begin{document}
\title{Supershells and the closest GRB remnants}

\author{J. E. Horvath\altaffilmark{1,2}}

\altaffiltext{1}{Steward Observatory, University of Arizona, 933 N. Cherry Av. , 85721 Tucson, AZ - USA}

\altaffiltext{2}{Instituto Astron\^omico e Geof\'\i sico
Universidade de S\~ao Paulo, Av. M.St\'efano 4200 - \'Agua Funda (04301-904) S\~ao Paulo SP - Brazil}

\begin{abstract}
Following recent suggestions of a gamma-ray burst origin of large
ISM holes (supershells) detected in the galaxy, we discuss the
effects of the recently discovered shell GSH 238+00+09 (and
other nearby extremely energetic supershells) on the
earth's atmosphere. We argue that the large
flux of gammas from these specific events
would have provoked major catastrophes to living
organisms. The lack of evidence for the latter can then
be used to rule out local isotropic  nearby GRBs
in the last tens of Myr and another origin for large supershells
is likely. Alternatively, this argument may be taken as
independent evidence for beaming of the gamma rays, although statistics are not yet
sufficient to constrain the beaming angle better than $\sim$ tens of degrees. The effects of
less-energetic events of the type of SN1998bw are also considered. It is shown that more than
a dozen of them should have occurred within $40 (E/10^{48} erg)^{1/2} \, pc$ of the earth in the last
500 Myr with similar effects. Beaming of the latter would also alleviate the conflict,
although another likely possibility (an actual rate much smaller than $10^{-4} yr^{-1}$)
may be a solution, at the same time avoiding an excessive number of supershells active in the galaxy.
\end{abstract}


\section{Introduction}

A breakthrough in GRB astrophysics has been achieved by the
observation of afterglows located $\sim $ few hours after the
event with unprecedented positional accuracy (\cite{cos97}). The
presence of absorption lines (\cite{met97,kul98}) in some of these
afterglows has convinced most researchers that most (if not all)
of the GRBs are extragalactic, although a through comprehension of
the bursts is still far away since the sources have yet to be
identified and the physics of the afterglows addressed (see, for
example, \cite{ree97,hur98}). Nevertheless, we may now assert that
a distance scale (and hence an energy scale) is available for
"classical" bursts. Regardless of the specific source, it is now
clear that the evidence points out to $E_{\gamma}$ as high as
$\simeq \, 10^{53} \, erg$ for the most energetic bursts
(\cite{kuk98}) if the gamma emission is isotropic.

While the study of distant, frequent bursts continues, the
observations have undoubtedly risen a number of questions related
to the occurrence of GRBs in the {\it local} universe. Thorsett
(1995) has discussed the effects of a close GRB on the earth's
biosphere (see also \cite{DLS98} for a related discussion). The
issue is timely since it has been shown that a burst must occur as
often as ($0.3-40$) $Myr$ per $L_{\ast}$, depending on the
evolution of the sources (\cite{wij98}). Loeb and Perna (1998)
have further suggested that most of the HI supershells could be
the remnants of GRBs. In fact the latest data on GRB afterglows
strongly suggest that some remnants must be found in a given
normal galaxy, since they should not dissipate before $\sim$ tens
of $Myr$. The two gigantic shells found by \cite{RV93} in NGC 4631
are perhaps the most clear examples that $\sim$ kpc-sized shells
requiring $\sim \, 10^{54} \, erg$ of input energy are real since
their identification is neater in external galaxies.

As discussed elsewhere (\cite{che74,hei79,ttb88}) the initial total energy of the expanding shell $E_{E}$
is inferred from the theory of explosions in a uniform medium. In the limit where the internal pressure is
no longer important, a numerical fit (\cite{che74}) yields
\begin{equation}
E_{E} \, = \, 1.2 \times 10^{53}
{\biggl( {n_{0} \over {cm^{-3}}} \biggr)}^{1.12}
{\biggl( {{R} \over {kpc}} \biggr)}^{3.12}
{\biggl( {v \over {km s^{-1}}} \biggr)}^{1.4} erg ,
\end{equation}
where $n_{0}$ is the particle number ambient density, $R$ is the radius of the shell and $v$ is the expansion velocity. The timescale
at which the supershell velocity is decelerated to $\simeq \, 8 \, km \, s^{-1}$
(the observed r.m.s. velocity of the interstellar clouds) is
$\tau \, = \, 0.31 \, R/v \, \simeq \, 37 \, (R/{1 \, kpc}) (v/8 \, km \, s^{-1})^{-1} $ Myr.
Since these estimates are derived from the simplest models (assuming spherical symmetry, uniform density, etc.);
there could be considerable
uncertainties when $E_{E}$ and $\tau$ are inferred from the observed data (see, e.g. \cite{hei79}).

While all the hitherto identified galactic supershells lie several
kpc away from the sun ($<D>$ $= \, 9.6 \, kpc$ for the Heiles 1979
compilation of expanding shells), the identification in HI, IR,
radio continuum and soft X-rays of a new, nearby supershell has
raised again the question of the relationship to GRBs and possible
biological effects. GSH 238+00+09 (\cite{hei98}) is not only a
major galactic supershell, but is also the only closer than $1 \,
kpc$. If, as argued by \cite{LP98} (see also \cite{BP98} and
\cite{EE98}), the supershell is the remnant of a galactic GRB we
may be have an unprecedented opportunity to correlate directly
cosmic explosions of this kind with fossil records.

\section{Gamma ray fluxes onto the earth and the ozone layer}

We have searched other large-energy shells in the catalogue of \cite{hei79} and found several candidates
(listed in Table 1) that, if identified with
remnants of GRBs, should have affected the earth in a very dramatic way (see below).
 It should be stressed that, mainly due to the difficulty of identifying
true large structures inside the galaxy, this selection is far
from definitive. For the sake of the argument and definiteness we shall address the objects in Table 1 only.

Consider the case of the simplest, "standard candle" scenario for GRBs. Given the geometrical centroids of the
supershells we may estimate immediately the flux of gamma-rays at the typical
passband 30-2000 keV at the top of the atmosphere $\phi$.
Table1 (column 6 ) displays these estimates for an assumed energy in gamma-rays of $E_{\gamma} \, = \, 10^{53} \, erg$.
Since the true luminosity distribution function is still an unsettled question and
there might be a considerable spread between the events, other possibilities should be considered. Since this is a
somewhat extreme assumption (although the continuing evidence for
$\sim \, 10^{53} \, erg$ from GRB 971214  lends some support to it,
see \cite{ode98}), we expect these numbers to be upper limits.

Thorsett (1995) pointed out that GRBs this close would
(because of the huge gamma fluxes of Table 1) have produced deep effects on the biosphere. The destruction
of a substantial amount
of the ozone layer along a $\sim \, 10 \, s$ typical burst duration
is the most obvious one, and seems inescapable since
the $\geq \, 10^{7} erg \, cm^{-2}$ gamma fluxes of Table1
are in fact larger than the equivalent total chemical energy of
the fragile ozone layer.

\placetable{tbl-1}

As discussed by \cite{SE95} (see \cite{rud74} for the first through discussion of a closely related event),
several general features of the incidence of a huge
gamma flux can be worked out with confidence. For example, it is well established that the production of large
concentrations of odd nitrogen $NO_{x}$ is  very harmful for the fragile ozone layer shielding the earth from solar
UV radiation. The dominating catalytic reactions are
\begin{equation}
NO \, + \, O_{3} \, \rightarrow \, NO_{2} \, + O_{2}
\end{equation}
\begin{equation}
NO_{2} \, + \, O \, \rightarrow \, NO \, + \, O_{3} \;,
\end{equation}
since their efficiency of ozone destruction is high. The additional $NO$ produced by the ionizing gamma flux will
greatly enhance the penetration of solar UV because the former is expected to be much higher than the steady
production by normal cosmic rays. The rate of production of $NO$ (in $mol/cm^{2}$) is
\begin{equation}
\xi \, = \, 10^{17} \phi_{7}
{\biggl[ {13\over{10 + y}}\biggr]} \;,
\end{equation}
where $\phi_{7} \equiv (\phi/10^{7} erg \, cm^{-2})$ is the
incident gamma flux scaled to a reference value, and the factor in
brackets is the ratio of efficiencies of the steady production to
the GRB flash in the stratosphere. Dividing $\xi$ by the
stratospheric column density and converting to parts per billion,
we derive the abundance of $NO$ produced by the GRB flash as the
physical solution of a quadratic equation, very well approximated
by
\begin{equation}
y_{flash} \, \simeq \, 51 \phi_{7}^{1/2} \, - \, 5 . \end{equation}
 Thus, the ratio of produced $[NO]$ to the present ambient $[NO]_{0}$
 is given by $X = (3+y_{flash})/3 \, \sim \, 16 \phi_{7}^{1/2}$.
Such a great abundance of $NO$ would remain in the stratosphere
for a mean residence time of $<\tau> \, = \, 4 \, yr$, which is
much larger than the homogenization time of the atmosphere. Thus,
once produced by the flash the ozone layer would be affected for a
period at least as large as the mean residence time of the
catalyzer.

The approximate formula employed by \cite{rud74} and \cite{SE95} to estimate that reduction is
\begin{equation}
{{[O_{3}]\over{[O_{3}]_{0}}}} \, = \,
{{(16 + 9 X^{2})^{1/2} - 3 X}\over{2}} \;,
\end{equation}
expected to be accurate to within a numerical factor. To be
conservative we shall regard as "catastrophic" those GRB producing
a flux $\phi$ which  kills at least $90 \%$ of the present $O_{3}$
layer through $NO$ enhancement (actually it is highly likely that
$\sim$ tens percent $O_{3}$ destruction would trigger massive
biological death ). Imposing that figure we obtain a lower bound
on $\phi$
\begin{equation}
\phi \, \geq \, 0.7 \times 10^{7} \, erg cm^{-2} \;,
\end{equation}
above which it is difficult  to envision survival of the species. This bound can be seen in Fig.1 ,
depicting at once the fraction of surviving $O_{3}$
as a function  of the incident flux $\phi_{7}$ . In other words , it is safe to state that all supershells
listed in Table 1 would have  trigger a massive extinction of life  through the ozone destruction . Thus ,
either the association is incorrect or beaming is necessarily involved in the emission.

\placefigure{fig-1}

\figcaption[]{The survival fraction of ozone ($[O_{3}]/[O_{3}]_{0}$ after the incidence of a
gamma flux $\phi$ (in $10^{7} \, erg \, cm^{-2}$ units). The hatched region corresponds to a destruction
fraction which would have caused a major extinction pattern. Compare with the values in the last column of Table 1}

In order to justify the assumed effects onto the biota we shall estimate, using a 
simple model, the killing timescale of a marine unicellular organism 
population exposed to the UVB (260-320 $nm$) radiation imediately after the burst.
For the bursts of Table 1 it is a good 
approximation to put that the afterburst solar UVB flux at sea level will be 
comparable to the one measured today at the top of the atmosphere 
$F_{\lambda} \, = \, 0.2 \, W \, cm^{-2} \, \mu m^{-1}$. The simplest 
theory of cell mortality by absorption of UV photons predicts the number of 
microorganisms to evolve according to

\begin{equation}
{N \over{N_{0}}} \, = \, 1 - {\bigl( 1 - \exp(- \kappa D)\bigr)}^{m}
\end{equation} 

where $D$ is the dose (here defined as $\int F_{\lambda} \, d \lambda$) and 
$m$ is the number of photons necessary to kill the cell. We shall apply this 
model to a marine population assumed to be  distributed exponentially with a 
depth scale $z_{0}$ (without day-night circulation) having a spontaneous 
reproduction rate $\eta$. If $N_{s}$ is the number of organisms at $z = z_{0}$ 
and the coefficient of attenuation for UVB photons in marine water is $z_{1}$ we 
find that the temporal evolution of this population at any depth will be given by 

\begin{equation}
N(z, t) = N_{s} \exp(-z/z_{0}) \, \exp[(\eta - \xi(z))t] 
\end{equation}

with $\xi(z) \, \simeq \, \kappa F_{\lambda 0} \, \Delta \lambda \exp(-z/z_{1})$. 
Now we may ask which is the time for killing $90 \%$ of these organisms once the 
UVB flux incides onto the sea surface, denoted as $\tau_{90}$. If we normalize the 
mortality curve using 
the data from modern bacteria (i.e. {\it Escherichia Coli}) we obtain for this 
time $\tau_{90} = 0.4 \exp(z/z_{1}) \, s$. Therefore, it is concluded that  
simple marine organisms, and especially those capable of photosynthesis, will be 
killed almost instantaneously unless they "hide" at several tens of $z_{1}$, in 
practice $\geq 100 m$ for a time as long as the healing of the ozone layer. 
Terrestrial organism behavior are much more difficult to model, although 
it has been long sice the '50 that mammals would not survive longer than 
$\sim \, 1 s$ without ozone. Even though simple models may be oversimplified, 
we believe that the essential points of a mass extinctions are adequately 
illustrated beyond any reasonable doubt.

The gamma shower would have produced other
unique catastrophes as well. The production of $\sim \, 10^{9} \, tons$ of
$NO_{x}$ enhancing the acid rains and the screening effects of $NO_{2}$
to the sunlight (with possible dramatic cooling effects, see \cite{rei78})
are just two of them. To address these issues properly,
the actual possibility of a close GRB calls for a
through study of the dynamical response of the biosphere to a large
perturbation, since all the effects are deeply interwoven and it is
quite difficult to isolate them due to their non-linear character.

\section{SN1998bw and supershells }

The reports (\cite{gal98,kul98}) on the remarkable type Ic supernova SN1998bw have prompted
an intense discussion in the literature on the nature of (some) core-collapse
events (see \cite{WES98,WW98}). The most unusual characteristics of this event are perhaps
the early, strong radio emission (\cite{kul98}), and the likely association (\cite{gal98}) with
GRB 980425. Striking as these features are, we here suggest that SN1998bw energetics may also hold a clue
for another long-standing puzzle of ISM structure and evolution, namely the
occurrence of supershells for which a growing body of observations (\cite{hei98,hei84} and references therein) is available.

Two separate groups (\cite{iwa98,WES98}) attempting
to model the light curve and observed spectra have shown that the event can be
understood as the spherically symmetric explosion of a massive star with the
ejection of $\geq \, 10 \, M_{\odot}$ of material including
$\sim \, 0.5 \, M_{\odot}$ of $^{56}$Ni. This gives in turn kinetic energies
of $\sim \, 3 \, \times \, 10^{52}$ erg (once the slightly different adopted
distances to the galaxy ESO 184-G82 are matched). Alternatively, an asymmetric
explosion has been advocated by \cite{HWW99}
which would in turn bring the $^{56}$Ni mass down to $0.2 \, M_{\odot}$ and the energy of the explosion to
$\sim \, 2 \, \times \, 10^{51}$ erg. While
asymmetric models share the attractive feature of being more akin to standard
core-collapse supernovae from the energetic and nucleosynthetic grounds, the
radio data gathered by \cite{kuk98} do not seem to show the
expected high polarization of a jet-like explosion. In fact, as already
mentioned by these authors, the low polarization is indeed consistent with
the simplest spherical blast wave models. While spherical models are in
serious trouble for producing a GRB, their energetics are much better tested
than the GRB production mechanisms, and are derived independently of the latter.
Therefore, it seems that
the existence of large, single stellar explosions
should be considered seriously (see \cite{iwa2000} for additional evidence of a second possible hypernova).

If we repeat  the calculations of the previous section and demand the same level
of ozone destruction to
trigger a biological catastrophe, we may derive instead an upper bound to the distance at which a "SN1998bw"
would be harmful. This distance is
\begin{equation}
D \, < \, 40 E_{48}^{1/2} \, pc \;,
\end{equation}
and we have scaled the energy in gamma rays to $E_{48} = (E/10^{48} erg)$. A consideration of the quotient of the
galactic disc to the volume defined by the distance eq.(8) gives us an idea of the frequency of the events
inside the latter if the galactic rate is known. Adopting a "SN1998bw" rate of
$\sim \, 10^{-4} yr^{-1}$, we obtain an effective rate for events closer than $D$
of $\simeq \, 3 \times 10^{-8} yr^{-1}$. Thus we expect $\sim \, 15$ events in
the last 500 Myr harmful for the ozone layer unless modest beaming is present. However, even for isotropically emitted gammas,
a much lower galactic rate would not only solve this problem but also reduce the number of active remnants to
tolerable levels, as pointed out above. To be sure, it is quite difficult to discriminate between
"classical GRB"-generated and "SN1998bw"-generated supershells (indeed, some of the objects in Table 1 could
belong to the latter instead of the former); but the conflict of the number of remnants will arise with a high
rate of $\sim \, 10^{-4} yr^{-1}$ simply because they live much longer in the ISM than their lower energy,
garden variety supernova cousins.

Precise determinations of the rate $\xi$ of explosion of such
energetic supernovae are not yet available, although it is
expected that a few of them should be present if $\xi \, > \,
\tau^{-1}$, where $\tau$ is the lifetime of the supershell.
Inserting the expression for $\tau $ given in section 2 we
conclude that a handful of  "active" energetic supernova-generated
shells if their rate is higher than $\sim \, 3 \, \times \,
10^{-8}$ yr$^{-1}$, which is almost two orders of magnitude lower
than the model-independent estimate (\cite{pac93}) for classical
GRBs and consistent with SN1998bw-type events being ten millon
times more rare than normal core-collapse supernovae (see
\cite{WW98}). \footnote{A very recent study by \cite{fra01} claims
a lower energy scale (and a correspondingly higher frequency of
the events by a factor $\sim \, 500$) that can be analyzed using
the results presented above.} Thus we think that there is ample
room for an actual galactic rate much lower than $10^{-4} yr^{-1}$
from this argument. By definition, supershells carry an energy of
at least tens of normal supernovae. The possible importance of the
SN1998bw event resides precisely in that its large inferred energy
comes to fill the gap between these two extremes, in spite of
widely different inferred gamma energies. It is indeed worthwhile
to note that the energy attributed to this supernova is within
$\sim \, 10 \%$ of the value required to produce the latest
identified (\cite{hei98}) supershell GSH 238+00+09 with a centroid
at only $\sim \, 0.8$ kpc away from the Sun. These estimates
suggest that very energetic supernovae, unfrequent as they might
be, are possibly an important component for ISM morphology and
evolution, with consequences for stellar formation and related
topics. Recent claims of detection in X-ray bands
\cite{WC99,wan99} need a follow-up to confirm the energetics and
to gather a statistically significative sample for the sake of
deeper analysis.

\section{Conclusions}

We have shown that the tentative identification of supershells
with GRB remnants leads to a definite conflict with biological
records, since the derived gamma fluxes onto the earth are
enormous and should then correlate with several massive
extinctions of life in the past $\sim \, 40 Myr$. Of course this
does {\it not} mean that some galactic GRB(s) could not have
affected the biology on earth \footnote {It is interesting (but
perhaps not significative) to note that GSH 139-03-69 should have
been almost simultaneous with the Priabonian extinction around 35
Myr ago where cool-temperature-intolerant organisms gradually
died, whereas GSH 242-03+37 has a characteristic age of 7.5 Myr
where even $^{10}Be$ marine sediments could be used for testing
purposes (\cite{mor91}).} , since epochs earlier than the lifetime
of the shells remain untested because it is impossible to find the
(now extinct) remnant supershell. If the supershells are indeed
GRB remnants and the gamma emission is beamed, the potential
conflict would vanish, although we estimate that the present
database of supershells would have to be enlarged by a factor of
$\sim  3$ to constrain the gamma emission to cones opening $\leq
\, 20$ degrees. It is likely that we will know the actual beaming by 
other methods much sooner.

The consideration of a lower energy (possibly associated with SN1998bw-like events) 
suggested also a modest
beaming effect (already discussed, for instance, by  \cite{HWW99} but from a completely 
different point of view
than ours) and/or the existence of a much lower rate to avoid overproduction of "hypernovae" 
remnants. A clue
to the energetics may be hidden in the discrepancy between the theoretical injection energy $E_{E}$
(\cite{iwa98,WES98}), amusingly sufficient to match exactly the energy derived for GSH 238+00+09, and the energy of
its associated GRB 980425 (\cite{SFP98})

We suggest that the preliminary consideration of the rate of these
SN-GRB type of events (\cite{WW98}) and their lifetimes is
nevertheless sufficient to consider them as serious candidates for
the progenitors of supershells. A closer examination of the latter
and modelling of the specific features of the former (as opposed
to the multiple-supernova-wind model) could prove useful to
reassess the formation and evolution of supershells. Clearly, much
work is needed before we pinpoint and understand the nature and
consequences of these events for the ISM and biological activity
with confidence.

\section{Acknowledgements}

We would like to acknowledge E. Reynoso for
useful advice on HI observations.
We are also grateful to E. M. G. Dal Pino and G.A Medina Tanco for encouragement and discussions on these subjects.
This work has been supported by FAPESP  Foundation
(S\~ao Paulo, Brazil). Steward Observatory colleagues and
staff are acknowledged for creating a stimulating working atmosphere during the realization of this work.


\clearpage

\begin{deluxetable}{crrrrr}
\footnotesize
\tablecaption{Features of selected supershells.See text for details\label{tbl-1}}
\tablewidth{0pt}
\tablehead{
\colhead{Shell GHS} & \colhead{$D$ (kpc)}    & \colhead{$v \, (km \, s^{-1}$)}
& \colhead{$\log E_{E}$}   & \colhead{$\tau \, (Myr)$} &
\colhead{$\phi (erg \, cm^{-2})$}
}

\startdata
238+00+09 & $\, \, 0.8$ & 9& 52.5& 21 & $1.4 \times 10^{9}$\nl
041+01+27 & $\, \, 2$ & 10 & 52.9 &7.4&$2.2 \times 10^{8}$\nl
075-01+39 & $\, \, 2.6$ & 22 & 52.9 &2.7& $1.3 \times 10^{8}$\nl
108-04-23 & $\, \, 2.5$ & 16 & 52.7 & 3& $1.4 \times 10^{8}$\nl
242-03+37 & $\, \, 3.6$ & 20 & 54.2& 7.4& $6.8 \times 10^{7}$\nl
061+00+51 & $\, \, 4.8$ & 14& 52.1& 33.5& $3.8 \times 10^{7}$\nl
016-01+71 & $\, \, 6.3$ & 18& 52.4& 20& $2.2 \times 10^{7}$\nl
139-03-69 & $\, \, 7.1$ & 18& 54.8 & 33& $1.7 \times 10^{7}$\nl
224+03+75 & $\, \, 7.6$ & 14& 53.4 & 13& $1.5 \times 10^{7}$\nl
029+00+133 & $\, \, 8.7$ & 20 & 52.6 & 6& $1.2 \times 10^{7}$\nl
\enddata

\end{deluxetable}
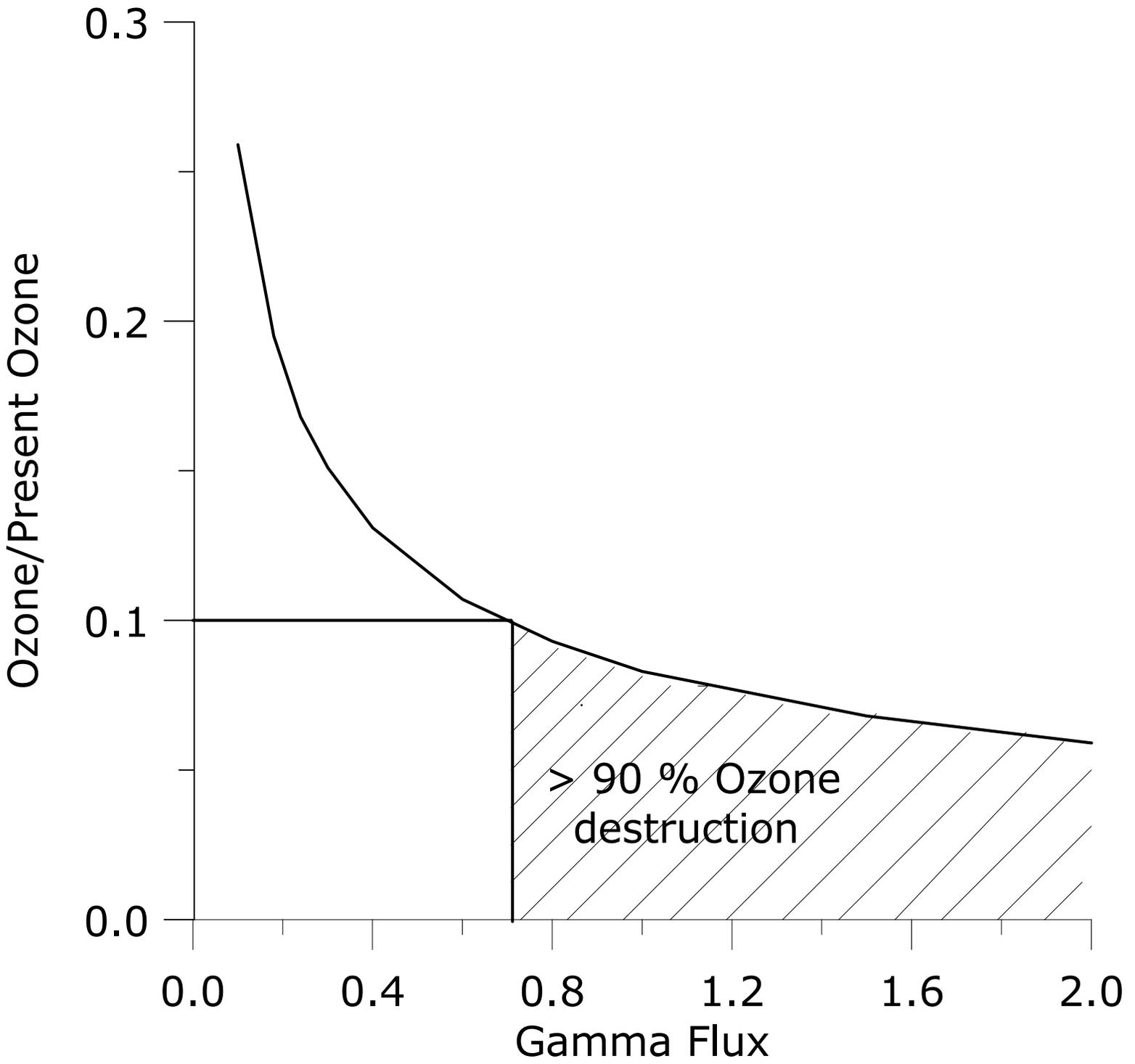
\end{document}